\documentclass[twocolumn,showpacs,prl,amsmath,amssymb]{revtex4}

\usepackage{graphicx}
\usepackage{bm}

\begin{document}

\title{Polaron percolation in diluted magnetic semiconductors}

\author{A. Kaminski}
\author{S. \surname{Das Sarma}}
\affiliation{
Department of Physics, University of Maryland, College Park, Maryland
20742-4111}

\begin{abstract}
We theoretically study the development of spontaneous magnetization in
diluted magnetic semiconductors as arising from a percolation of bound
magnetic polarons. Within the framework of a generalized percolation
theory we derive analytic expressions for the Curie temperature and
the magnetization, obtaining excellent quantitative agreement with
Monte Carlo simulation results and good qualitative agreement with
experimental results.
\end{abstract}


\maketitle

Diluted magnetic semiconductors (\emph{e. g.} In$_{1-x}$Mn$_x$As,
Ga$_{1-x}$Mn$_x$As), which are materials with some fraction of the
nonmagnetic lattice atoms replaced by magnetic atoms (\emph{i. e.}  a
fraction $x$ of Ga or In atoms being randomly replaced by Mn atoms),
have attracted a great deal of attention
\cite{sds1,sds3,sds7,sds8,sds2,sds10,sds12,sds6,sds9,sds11} following
the recent discovery \cite{sds4} of ferromagnetism in
Ga$_{1-x}$Mn$_x$As, with a Curie temperature $T_c\approx 100$K in the
$x=0.03-0.07$ range. The subject is of considerable fundamental and
technological interest.  Technologically, a semiconductor, which is
also a ferromagnet, raises the exciting potential of spintronic
applications \cite{sds5} where logic and memory operations could in
principle be seamlessly integrated on a single device. From a
fundamental perspective, understanding ferromagnetism in a novel
material (which is also a semiconductor) is an important challenge,
particularly so in this case in light of the extensive existing
earlier work involving doped II-VI semiconductor materials (as against
the systems of current interest, which are magnetically doped III-V
systems) which were never convincingly found to be ferromagnetic
except perhaps with extremely low values of $T_c$. It is therefore not
surprising that there has been a great deal of theoretical activity
\cite{sds1,sds3,sds7,sds8,sds2,sds10,sds12,sds6,sds9,sds11} trying to
understand the ferromagnetic mechanism in GaMnAs. Although no
theoretical consensus on the precise ferromagnetic mechanism has yet
been reached in the literature it is now established that the
interaction between the magnetic atoms, which leads to the
ferromagnetic phase at low enough temperatures, is induced by charge
carriers (holes in the case of GaMnAs) in the semiconductor host.
Theoretical investigation of these systems is hampered by the fact
that both disorder and interactions are strong and must be taken into
account non-perturbatively. This problem has been approached in a
number of ways, including theoretical approximations assuming charge
carriers being almost free~\cite{sds1,sds3,sds7,sds8} and numerical
studies in the opposite limit of strongly localized charge
carriers~\cite{sds2,sds10,sds12}. However, a comprehensive
understanding of the physics of diluted magnetic semiconductors has
not been achieved yet.

Theoretical approaches \cite{sds1,sds3,sds7,sds8} treating the charge
carriers as free carriers in the valence band of the semiconductor
have employed the simple Weiss mean field theory incorporating the
band structure details (\emph{e.g.}  spin-orbit coupling, etc.),
without taking into account effects of disorder. While this approach
is claimed to provide quantitatively accurate values of $T_c$, it
fails qualitatively in accounting for experimentally observed
transport properties \cite{sds13} of GaMnAs where the resistivity is
very high and seems to obey a Mott variable range hopping behavior at
low temperatures reminiscent of an insulating system.  In addition,
the resistivity always decreases with increasing temperature above
$T_c$, which is again typical of a localized insulating system.  The
free-carrier mean-field approach also fails to explain the recently
reported \cite{sds14} dependence of $T_c$ on annealing which indicates
a crucial role for disorder.  A recent dynamical mean field theory
calculation \cite{sds11} concluded that the ferromagnetic GaMnAs may
actually belong to a situation where the carriers are just on the edge
of being strongly localized.  The approach \cite{sds2,sds10,sds12}
treating the carriers as localized carriers in a semiconductor
impurity band (thus being opposite to the the free-carrier mean-field
approach), which has so far been explored only numerically, also leads
to reasonable agreement with experiments and indicates a very strong
dependence of $T_c$ on disorder.  In this paper we provide an
analytical theory which takes into account both disorder and strong
magnetic interaction starting with localized carriers and using the
physically appealing magnetic polaron percolation picture.  Our theory
has a starting point similar to that in Refs.~\cite{sds2,sds10,sds12}
except that ours is a completely analytical physical theory in
contrast to the numerical approach used in the works
\cite{sds2,sds10,sds12}. Where applicable our analytical results agree
well with the numerical results of Refs.~\cite{sds2,sds10,sds12}.

\begin{figure}
\includegraphics{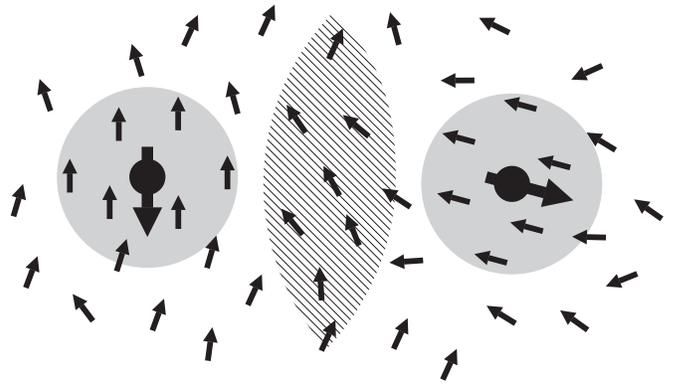}
\caption{\label{fig:twopolarons} Interaction of two bound magnetic
  polarons (after Ref.\protect\cite{WolffEtAl96}) The polarons are
  shown with gray circles, small and large arrows show impurity and
  hole spins respectively.}
\end{figure}

In this paper we consider a system in which transition to the
insulating state due to localization of the charge carriers occurs at
temperatures higher than the Curie temperature $T_c$.  The carriers
will be called ``holes'' throughout the text of this paper since in
GaMnAs the carries are holes although the theory to be developed in
this paper applies equally to the situation where the carriers are
electrons.  Exchange interaction of localized holes with magnetic
impurities leads to the formation of bound magnetic polarons
\cite{sds15,WolffEtAl96}.  Since the concentration of magnetic
impurities is much larger than the hole
concentration~\cite{MatsukuraEtal98}, most likely due to compensation
by As antisite defects, a bound magnetic polaron consists of one
localized hole, and a large number of magnetic impurities around the
hole localization center. Even though the direct exchange interaction
of the localized holes is antiferromagnetic, the interaction between
bound magnetic polarons may be ferromagnetic \cite{WolffEtAl96} at
large enough concentrations of magnetic impurities. To understand the
physics of this phenomenon, one may consider two neighboring polarons
(Fig.~\ref{fig:twopolarons}).  The localized holes of these polarons
both act on the impurities surrounding them thus producing an
effective magnetic field for these impurities. The energy minimum is
reached by this system when the impurity spins are parallel to this
effective field, and the magnitude of the field is maximum.  The
maximum of this effective magnetic field is achieved when the spins of
the localized holes are parallel.  Therefore at low temperatures the
system should eventually reach the state where the spins of all holes
point in the same direction, and all impurity spins point in the same
or in the opposite direction, depending on the sign of the
impurity-hole exchange interaction.

If the hole localization radius is much less than the characteristic
distance between the localized holes, the disorder in the hole
positions must have dramatic effect on the whole picture of the
ferromagnetic transition. This notion has been confirmed by Berciu and
Bhatt \cite{sds10}, who have shown, by means of numerical simulations,
that both the Curie temperature and the shape of the magnetization
curve $M(T)$ are indeed strongly affected by disorder.  It has been
known that the percolation theory \cite{EfrosShklovskiiBook} provides
many adequate tools to deal with ferromagnetism in disordered systems
with strong localization of carriers \cite{KorenblitShender78}. In
this paper we present a quantitative description of the spontaneous
magnetization in magnetic semiconductors within the framework of the
percolation theory.

In our model, the charge carriers are localized.  The hole wave
function is assumed to fall off exponentially away from localization
centers, with decay length $a_B$.  We consider the low carrier density
regime in which the mean distance between the localized holes is much
larger than the hole localization radius, $a_B^3 n_h\ll 1$. We note
that for GaMnAs, $a_B\approx 10$ \AA, and therefore our theory applies
in the regime $n_h\ll 10^{21}\ \textrm{cm}^{-3}$, with experimental $n_h$
values currently being around $10^{19}\ \textrm{cm}^{-3}$. The localization
centers are distributed randomly in the sample.  Magnetic impurities
are distributed within the sample with concentration $n_i\gg n_h$
randomly as well. The Hamiltonian of the system has the form
\begin{equation}
\label{H1}
\hat{H}= \sum_{kj}J_{kj}\hat{\mathbf{S}}_k
\hat{\mathbf{s}}_j,
\end{equation}
where indices $k$ and $j$ label magnetic impurities and holes
respectively, $\hat{\mathbf{S}}_k$/$\hat{\mathbf{s}}_j$ are the
impurity/hole spin operators. Matrix elements $J_{kj}$ of the
impurity-hole exchange interaction decay exponentially with the
distance between the interacting impurity and hole,
$J_{kj}=J_0\exp(-2|\mathbf{r}_k-\mathbf{r}_j|/a_B)$. The direct
(antiferromagnetic) exchange interaction between the magnetic
impurities is neglected, since their relative concentration $x$ in the
lattice of the host semiconductor is much less than unity although
this may be important for larger values of $x$, leading to the
suppression of $T_c$.  We will not discuss possible mechanisms of hole
localization, since the properties of our model hold for any of them
as long as the decay of the localized hole's wave function is
exponential.

At some temperature $T$, magnetic impurities that are at distances
$r<R_p(T)\equiv (a_B/2)\ln(sS|J_0|/T)$ from hole localization centers
have their spins strongly correlated with the spins of the
corresponding holes (here $s$ and $S$ are the absolute values of the
hole or impurity spin respectively).  The spins which do not have a
localized hole within a circle of radius $R_p(T)$ around them are
essentially free.  The quantity $R_p(T)$ is the effective radius of a
magnetic polaron; it grows as the temperature is lowered.  Clearly at
low enough temperatures neighboring magnetic polarons overlap and
interact with each other via interaction with impurities between
them~\cite{WolffEtAl96}.  This interaction produces alignment of the
polaron spins. When the cluster of correlated polarons having the size
of the sample (the infinite cluster) appears, the ferromagnetic
transition occurs.

Even though the picture presented above is qualitatively correct, one
needs to find the maximum characteristic temperature
$T_{2p}(r)$ at which the spins of two polarons at a given distance $r$
from each other are still strongly correlated. The Hamiltonian of a
two-polaron subsystem is given by Eq.~(\ref{H1}), where hole index $j$
takes only two values $j_1$ and $j_2$ corresponding to the two
polarons under consideration. Our present goal is to find the
characteristic spin-correlation temperature $T_{2p}(r)$ at given $r$,
$a_B$, $J_0$, and $n_i$. Since we are interested in the system
behavior only at and below the percolation transition, we may limit
our consideration to polaron pairs with $r\gtrsim n_h^{-1/3}\gg a_B$.
A rough preliminary estimate, which can be obtained from
straightforward dimensional analysis, reads
\begin{equation}
T_{2p}(r)\sim A(r, a_B, n_i)s S |J_0| \exp\left(-r/a_B\right)\;,
\label{T2pestimate}
\end{equation}
where dependence of the dimensionless prefactor function $A(r, a_B,
n_i)$ on its arguments is weaker than exponential. 
Since at $T\sim sS|J_0|\exp(-r/a_B)$ the two polarons \emph{must} be
correlated, $A \ge 1$.  At temperatures of the order of the right-hand
side of Eq.~(\ref{T2pestimate}), each polaron already has a large
number of impurity spins near its center strongly polarized along the
direction of the spin of its hole. Due to this coupling, the spin of
the polaron's hole becomes ``massive,'' and may be considered a
classical field as far as its interaction with more remote impurities
is concerned. This effective field produces polarization of the remote
impurities; the characteristic value of the projection $S_k^{(j_1)}$
of the spin of $k$th impurity onto the spin direction of the $j_1$-th
polaron is roughly $S \min\{1,s J_{kj_1}/T\}$.  The characteristic
energy of interaction of this impurity with the other polaron is of
the order of $s S J_{kj_2}$, therefore the contribution of the $k$th
impurity to the interaction of two polarons is of the order of $S^2 s
J_{kj_2} \min\{1,s J_{kj_1}/T\}$ [here we assume that $sS J_{kj_1}, T
> sSJ_{kj_2}$ since $T_{2p}(r) \ge s S |J_0| \exp(-r/a_B)$].  This
estimate allows us to determine which impurities are important for the
interaction of two given polarons. Namely, these impurities are the
ones in the lens-shaped region between the two localization centers,
see Fig.~\ref{fig:twopolarons}. The diameter of this region is of the
order of $\sqrt{r a_B}$; more remote impurities are interacting too
weakly with both polarons to be of any importance. Finding the
thickness of the lens-shaped region is less trivial. As we consider
impurities away from the middle point and closer, say, to $j_1$,
decrease of $J_{kj_2}$ is fully compensated by the increase of
$J_{kj_1}$, until we enter the region where the spins of the
impurities are saturated due to their proximity to $j_1$. Thus the
size of the region of interest in the direction along
$\bm{r}_{j_1}-\bm{r}_{j_2}$ depends on the temperature. Since the
characteristic correlation temperature $T_{2p}(r)$ is the quantity to
find, we will proceed by iterations, starting with the value given by
the right-hand side of Eq.~(\ref{T2pestimate}) with $A(r, a_B, n_i) =
1$ (so the thickness of the lens-shaped interaction region equals
$a_B$). As we will see, just one iteration yields the answer with good
precision.

We will take into account only the impurities in the interaction
region; the total number of these impurities equals $N\sim a_B^2 r
n_i$ in the first iteration. The coupling of each of these impurities
to either hole is approximately $J(r/2)\equiv J_0\exp(-r/a_B)$. It
allows us to reduce the Hamiltonian to the following form:
\begin{equation}
\hat{H}=s J(r/2) \cos\frac{\theta}{2} \sum_{k=1}^N\bm{S}_k^{(z)}\;.
\label{H2peff}
\end{equation}
Here the angle between the spins of the two polarons is denoted
$\theta$, and the direction of the $z$ axis is chosen along the
direction of the vector $\bm{s}_{j_1} + \bm{s}_{j_2}$. We have
neglected spatial variation of the polaron exchange field in the
interaction region, since the relative magnitude of this variation
does not exceed unity.

The partition function of the system described by Hamiltonian
(\ref{H2peff}) can easily be found, and after straightforward
algebra we arrive at the expression for the average cosine of the
angle $\theta$ between the spins of the two interacting polarons:
\begin{equation}
\label{avcosresult}
\langle\cos\theta\rangle\sim\left\{
\begin{array}{ll}
N [sSJ(r/2)/T]^2\;, & \sqrt{N}sS|J(r/2)|\ll T\\
1- \frac{1}{N[sSJ(r/2)/T]^{2}}\;, & \sqrt{N}sS|J(r/2)|\gg T \;.
\end{array}
\right.
\end{equation}
Therefore, the spins of the two polarons at distance $r$ are
correlated at temperatures below $\sqrt{N}sS|J(r/2)|$. Using the estimate
$N\approx a_B^2 r n_i$, we finally
arrive at
\begin{equation}
\label{T2pfinal}
T_{2p}(r)\sim a_B \sqrt{r n_i} sS |J_0| \exp\left(-r/a_B\right)\;.
\end{equation}
The maximum possible distance $r_{corr}(T)$ between two polarons with
correlated spins at a given temperature $T$ is therefore
\begin{equation}
\label{rcorr}
r_{corr}(T) \sim a_B \left[ \ln\frac{sS|J_0|}{T} +
\frac12\ln\left(a_B^3 n_i\ln\frac{sS|J_0|}{T}\right)
\right]\;.
\end{equation}
Now we can use result (\ref{T2pfinal}) to get a better estimate for
the thickness of the lens-shaped interaction region in order to
perform the next iteration. It is determined by the condition
$J_{kj_{1,2}} < T$, so the thickness is of the order of $a_B \ln (a_B
\sqrt{r n_i})$. The resulting correction to $T_{2p}(r)$ is just a
factor of $\ln (a_B \sqrt{r n_i})$. Taking this correction into
account is clearly beyond the accuracy limits of the approximation
used above, so we neglect it, and use Eq.~(\ref{T2pfinal}) as our
final result.

Now we consider a system of randomly placed magnetic polarons. As the
temperature is being lowered, the spins of neighboring polarons become
aligned, and clusters of polarons with the same spin appear. At any
given temperature $T$, the polarons separated by a distance smaller
than $r_{corr}(T)$ are joined into magnetic clusters. The lower the
temperature, the more such ``links'' between polarons are established,
and the larger the average cluster size.  Finally, at low enough
temperatures, a cluster having the dimensions of the sample, the so
called ``infinite cluster'' appears, and the magnetization of the
sample acquires some finite value. The problem of finding the
transition temperature is identical to the problem of finding the
critical percolation radius in the problem of randomly placed
overlapping spheres~\cite{EfrosShklovskiiBook}. The latter problem has
been solved numerically, and it has been demonstrated that the
infinite cluster forms in the system when the link length reaches the
value $r_{perc} \approx 0.86/\sqrt[3]{n_h}$~\cite{PikeSeager74}.
Substituting this expression into Eq.~(\ref{T2pfinal}), we get the
expression for the ferromagnetic transition temperature
\begin{equation}
\label{Tc}
T_c\sim a_B \sqrt{n_i} n_h^{-1/6}sS|J_0| 
\exp\left(-\frac{0.86}{a_B\sqrt[3]{n_h}}\right)\;.
\end{equation}
The limit of
applicability of Eq.~(\ref{Tc}) is determined by the condition $a_B^3
n_h \ll 1$. A similar exponent was obtained in
Ref.~\cite{KorenblitEtAl73} for ferromagnetic transition in PdFe
alloys.

It is instructive to point out that our Eq.~(\ref{Tc}) is consistent
with the mean-field result derived in the literature
\cite{sds1,sds3,sds7,sds8} for the opposite limit of almost free
holes,
\begin{displaymath}
\label{Tcmf}
T_c \sim n_i  J_{\textrm{mf}}^2 \xi(T_c)\;,
\end{displaymath}
where $J_{\textrm{mf}}$ is related to $J_0$ of Eq.~(\ref{H1}) by
$J_{\textrm{mf}}\sim a_B^3 J_0$, and $\xi(T)\sim
n_h/\max\{\varepsilon_F, T\}$ is the magnetic susceptibility of holes,
with $\varepsilon_F$ being the Fermi energy. In the case of low
electron density, $T\gg \varepsilon_F$
\begin{displaymath}
T_c \sim \sqrt{n_in_h} |J_{\textrm{mf}}|\;,
\end{displaymath}
which matches the result (\ref{Tc}) of the bound-polaron picture at
the limit of applicability of the latter, $a_B^3 n_h \sim 1$.

Since $n_h \ll n_i$, one polaron includes many magnetic impurities,
and the total magnetization of the sample is that of impurities:
\begin{equation}
\label{magn}
M(T) = S n_i {\cal V}[ r_{corr}(T, n_i, n_h, a_B) \sqrt[3]{n_h} ]\;,
\end{equation}
where $r_{corr}$ is defined by Eq.~(\ref{rcorr}), the and universal
function ${\cal V}(y)$ is the infinite cluster's volume in the model
of overlapping spheres; it depends only on the product $y$ of the
spheres' diameter and the cubic root of their concentration.

\begin{figure}
\includegraphics{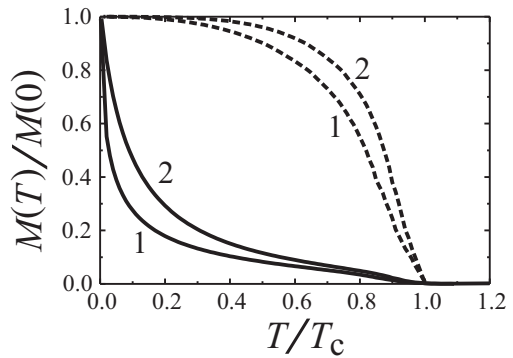}
\caption{\label{fig:graphs} The solid lines show the
  relative magnetization of the magnetic impurities
  [Eq.~(\protect\ref{magnfinal})] for $a_B^3 n_h = 5\times 10^{-3}$
  (curve 1) and $2\times 10^{-2}$ (curve 2). The dashed lines show the
  relative magnetization of localized holes, whose contribution to the
  total sample magnetization is small. }
\end{figure}

Using Eqs.~(\ref{rcorr}) and (\ref{Tc}), we cast Eq.~(\ref{magn}) in
the form
\begin{equation}
\label{magnfinal}
\frac{M(T)}{M(0)}= 
{\cal V}\left(0.86+\left(a_B^3 n_h\right)^{\frac{1}{3}}
\ln \frac{T_c}{T}\right),
\end{equation}
with the Curie temperature $T_c$ given by Eq.~(\ref{Tc}).  One can see
that the shape of the magnetization curve is determined by only one
dimensionless parameter $a_B^3 n_h$, while the expression (\ref{Tc})
for $T_c$ is more complicated and depends on all parameters of the
model. Fig.~\ref{fig:graphs} shows the temperature dependence of the
magnetization at two values of $a_B^3 n_h$; the curve is more concave
at smaller values of this parameter, which is precisely the
experimental observation. Our magnetization results agree with the
numerical results of Ref.~\cite{sds10} and are consistent with the
experimental magnetization data in GaMnAs, particularly for systems
with lower values of $T_c$ where our polaron percolation picture
applies better due to stronger carrier localization associated with
lower values of $a_B^3 n_h$.

To conclude we have developed an analytic polaron percolation theory
for DMS ferromagnetism in the limit of low carrier density or
equivalently strong carrier localization $a_B^3 n_h \ll 1$.
Interestingly our polaron percolation theory reproduces the free
carrier Weiss mean field theory in the limit of $a_B^3 n_h \sim 1$.
Our analytic results are in good agreement with existing numerical
results in the strongly localized limit \cite{sds2,sds10,sds12}. The
experimental DMS currently have $a_B^3 n_h\sim 10^{-1} - 10^{-3} $
which makes our theory marginally applicable to the experimental
systems, and we get reasonable agreement with experimental results for
$T_c$ and for $M(T)$.

This work was supported by the US-ONR, the LPS, and DARPA. We thank B.
I.  Shklovskii for useful discussion on percolation theory.

\end{document}